\documentclass[twocolumn,showpacs,showkeys,reprint,amsmath,amssymb,aps]{revtex4-1}

\usepackage{graphicx}% Include figure files
\usepackage{dcolumn}% Align table columns on decimal point
\usepackage{bm}% bold math
\usepackage{tabularx}% Table width
\usepackage{amsmath}% Higher math
\usepackage{xcolor}

\begin{document}

\title{Competing superconductivity and charge-density wave in Kagome metal CsV$_3$Sb$_5$: evidence from their evolutions with sample thickness}

\author{B. Q. Song,$^{1}$ X. M. Kong,$^{1}$ W. Xia,$^{2,3}$ Q. W. Yin,$^4$ C. P. Tu,$^{1}$ C. C. Zhao,$^{1}$ D. Z. Dai,$^{1}$ K. Meng,$^{1}$ Z. C. Tao,$^2$ Z. J. Tu,$^4$ C. S. Gong,$^4$ H. C. Lei,$^{4,\dag}$ Y. F. Guo,$^{2,\ddag}$ X. F. Yang,$^{1,\sharp}$ and S. Y. Li$^{1,5,6*}$}

\affiliation
 {$^1$State Key Laboratory of Surface Physics, Department of Physics, Fudan University, Shanghai 200438, China\\
 $^2$School of Physical Science and Technology, ShanghaiTech University, Shanghai 201210, China\\
 $^3$ShanghaiTech Laboratory for Topological Physics, Shanghai 201210, China\\
 $^4$Department of Physics and Beijing Key Laboratory of Opto-electronic Functional Materials and Micro-nano Devices, Renmin University of China, Beijing 100872, China\\
 $^5$Collaborative Innovation Center of Advanced Microstructures, Nanjing 210093, China\\
 $^6$Shanghai Research Center for Quantum Sciences, Shanghai 201315, China
}

\date{\today}

\begin{abstract}
Recently superconductivity and topological charge-density wave (CDW) were discovered in the Kagome metals $A$V$_3$Sb$_5$ ($A$ = Cs, Rb, and K), which have an ideal Kagome lattice of vanadium. Here we report resistance measurements on thin flakes of CsV$_3$Sb$_5$ to investigate the evolution of superconductivity and CDW with sample thickness. The CDW transition temperature ${\it T}_{\rm CDW}$ decreases from 94 K in bulk to a minimum of 82 K at thickness of 60 nm, then increases to 120 K as the thickness is further reduced to 4.8 nm (about five monolayers). Since the CDW order in CsV$_3$Sb$_5$ is quite three-dimensional (3D) in the bulk sample, the non-monotonic evolution of ${\it T}_{\rm CDW}$ with reducing sample thickness can be explained by a 3D to 2D crossover around 60 nm. Strikingly, the superconducting transition temperature ${\it T}_{\rm c}$ shows an exactly opposite evolution, increasing from 3.64 K in the bulk to a maximum of 4.28 K at thickness of 60 nm, then decreasing to 0.76 K at 4.8 nm. Such exactly opposite evolutions provide strong evidence for competing superconductivity and CDW, which helps us to understand these exotic phases in $A$V$_3$Sb$_5$ Kagome metals.
\end{abstract}

\maketitle

The recently discovered V-based Kagome metals, $A$V$_3$Sb$_5$ ($A$ = Cs, Rb, and K), have stimulated great interest in the field of condensed matter physics \cite{Rb-K-Cs,KVSb-Z2,RbVSb-SC,CsV3Sb5-Z2}. These compounds, consisting ideal Kagome lattice of vanadium coordinated by antimony, show superconductivity with superconducting transition temperatures (${\it T}_{\rm c}$) of 2.5, 0.92, and 0.93 K for $A$ = Cs, Rb, and K, respectively \cite{CsV3Sb5-Z2,KVSb-Z2,RbVSb-SC}. Besides superconductivity, CDW transitions are revealed in the normal states at ${\it T}_{\rm CDW}$  78\textendash104 K for $A$V$_3$Sb$_5$ by X-ray Diffraction (XRD) and scanning tunnelling microscopy (STM) measurements \cite{CsV3Sb5-Z2,KVSb-Z2,RbVSb-SC,topological_charge_order_KVSb}. The angle-resolved photoemission spectroscopy (ARPES) experiemnts point out a topological surface state with multiple Dirac nodal points close to the Fermi level, suggesting that $A$V$_3$Sb$_5$ can be categorized as topological Kagome metals \cite{Rb-K-Cs,CsV3Sb5-Z2,KVSb-Z2}.

Furthermore, the STM study revealed a topological charge order in KV$_3$Sb$_5$ with chiral anisotropy \cite{topological_charge_order_KVSb}, which may cause the giant anomalous Hall effect and a possibility of unconventional superconductivity \cite{topological_charge_order_KVSb,anomalous Hall effect}. Indeed, nodal superconductivity and a pressure-induced double dome superconductivity were found by the ultralow-temperature thermal conductivity and high-pressure resistance measurements in CsV$_3$Sb$_5$ \cite{SYLiCsVSb}. Unconventional strong-coupling superconductivity in these V-based superconductors was also suggested by Josephson STM \cite{Strong-coupling}. Although the penetration depth and nuclear magnetic resonance (NMR) measurements claimed $s$-wave superconductivity \cite{Nodeless superconductivity,s-wave superconductivity}, the subsequent ultralow-temperature STM demonstrated both nodal and nodaless gaps for CsV$_3$Sb$_5$ with multiple Fermi surfaces \cite{Multiband superconductivity,Fermi surface mapping}. The exact locations of the gap nodes still need to be identified. In this context, the Kagome metal $A$V$_3$Sb$_5$ provides a great platform to study the interplay of superconductivity, CDW, and topological band structure.

Previously, the superconducting dome under low pressure has already indicated the competition between superconductivity and CDW in $A$V$_3$Sb$_5$ \cite{SYLiCsVSb,J.-G. Cheng,Z. Yang,X. Chen,Pressure-KVSb}. In this Letter, we investigate the evolution of superconductivity and CDW with sample thickness in the mechanically exfoliated CsV$_3$Sb$_5$ thin flakes by electrical transport measurement. Two non-monotonic evolutions are revealed, and the one for CDW can be explained by a 3D to 2D crossover around 60 nm. It is striking that the two evolutions are exactly opposite for superconductivity and CDW, demonstrating the competition between them in CsV$_3$Sb$_5$.

Single crystals of CsV$_3$Sb$_5$ were grown by the self-flux method \cite{CsV3Sb5-Z2,SYLiCsVSb}. Thin flakes of CsV$_3$Sb$_5$ were prepared by an Al$_2$O$_3$-assisted exfoliation method \cite{Al2O3}. Al$_2$O$_3$ thin film with thickness ranging from 60 to 100 nm was deposited by thermally evaporating Al under an oxygen pressure of 10$^{-2}$ Pa on the freshly prepared surface of the bulk CsV$_3$Sb$_5$ single crystal. Then the Al$_2$O$_3$ film was picked up with a thermal release tape, along with pieces of CsV$_3$Sb$_5$ thin flakes separated from the bulk. The Al$_2$O$_3$/CsV$_3$Sb$_5$ stack was subsequently released onto a piece of polydimethylsiloxane (PDMS) upon heating, with the CsV$_3$Sb$_5$ side in contact with the PDMS surface. Next the PDMS was stamped onto a substrate and was peeled away, leaving the Al$_2$O$_3$ film covered with CsV$_3$Sb$_5$ thin flakes on the substrate. Figure 1(b) displays an optical image of CsV$_3$Sb$_5$ thin flake on Al$_2$O$_3$ film supported on a 300 nm SiO$_2$/Si substrate. Then the electrodes were fabricated on the CsV$_3$Sb$_5$ thin flake with direct metal deposition through stencil masks. All the devices were fabricated in an argon atmosphere with O$_2$ and H$_2$O content kept below 0.5 parts per million to avoid sample degradation.

The thickness of the CsV$_3$Sb$_5$ thin flakes was determined by Atomic Force Microscopy (AFM) (Park NX10). The resistance measurements of the thin flakes were performed in a physical properties measurement system (PPMS, Quantum Design) and a $^3$He refrigerator. All the loading processes out of the glove box were done within one minute to prevent sample from degradation.

\begin{figure}
\includegraphics[clip,width=9cm]{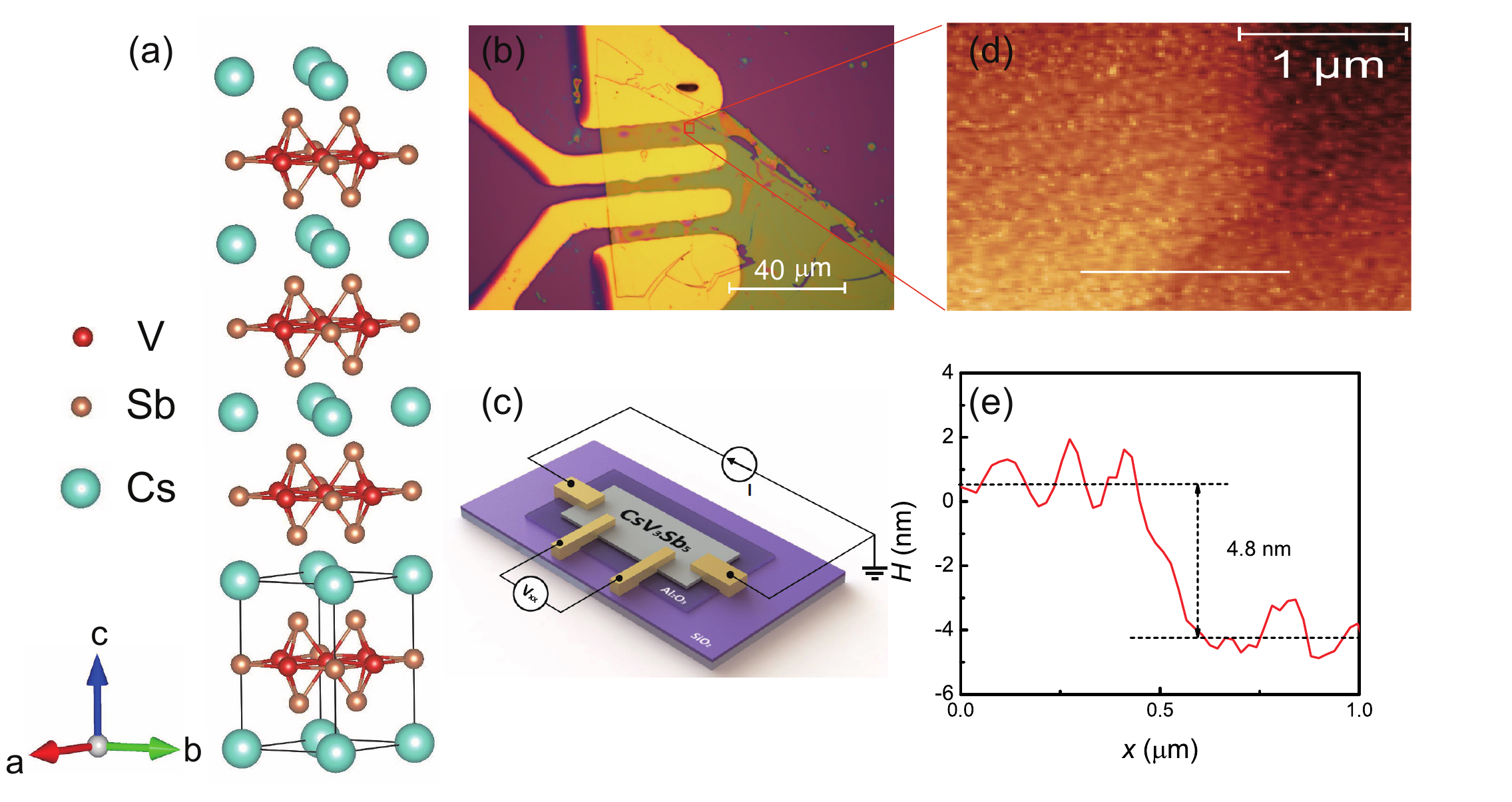}
\caption{(a) Crystal structure of CsV$_3$Sb$_5$. The layers consisting ideal Kagome lattice of vanadium coordinated by antimony are separated by the alkali-metal atoms of Cs. The lattice parameters are $a$ = $b$ = 5.4949(3) \AA\ and $c$ = 9.3085(5) \AA\ \cite{Rb-K-Cs}. (b) Optical image of a thin CsV$_3$Sb$_5$ flake device. Cr/Au contacts were deposited on the sample through the stencil mask for four-probe measurements. The scale bar is 40 $\mu$m. (c) Schematic structure of CsV$_3$Sb$_5$ device and measurement set-up. (d) Atomic force microscopy image of the area marked by the square in (b). The scale bar is 1 $\mu$m.  (e) Cross-sectional profile of the thin flake along the white line in (d). The step is about 4.8 nm in height which corresponds to five CsV$_3$Sb$_5$ monolayers.}
\end{figure}

\begin{figure}
\includegraphics[clip,width=8.38cm]{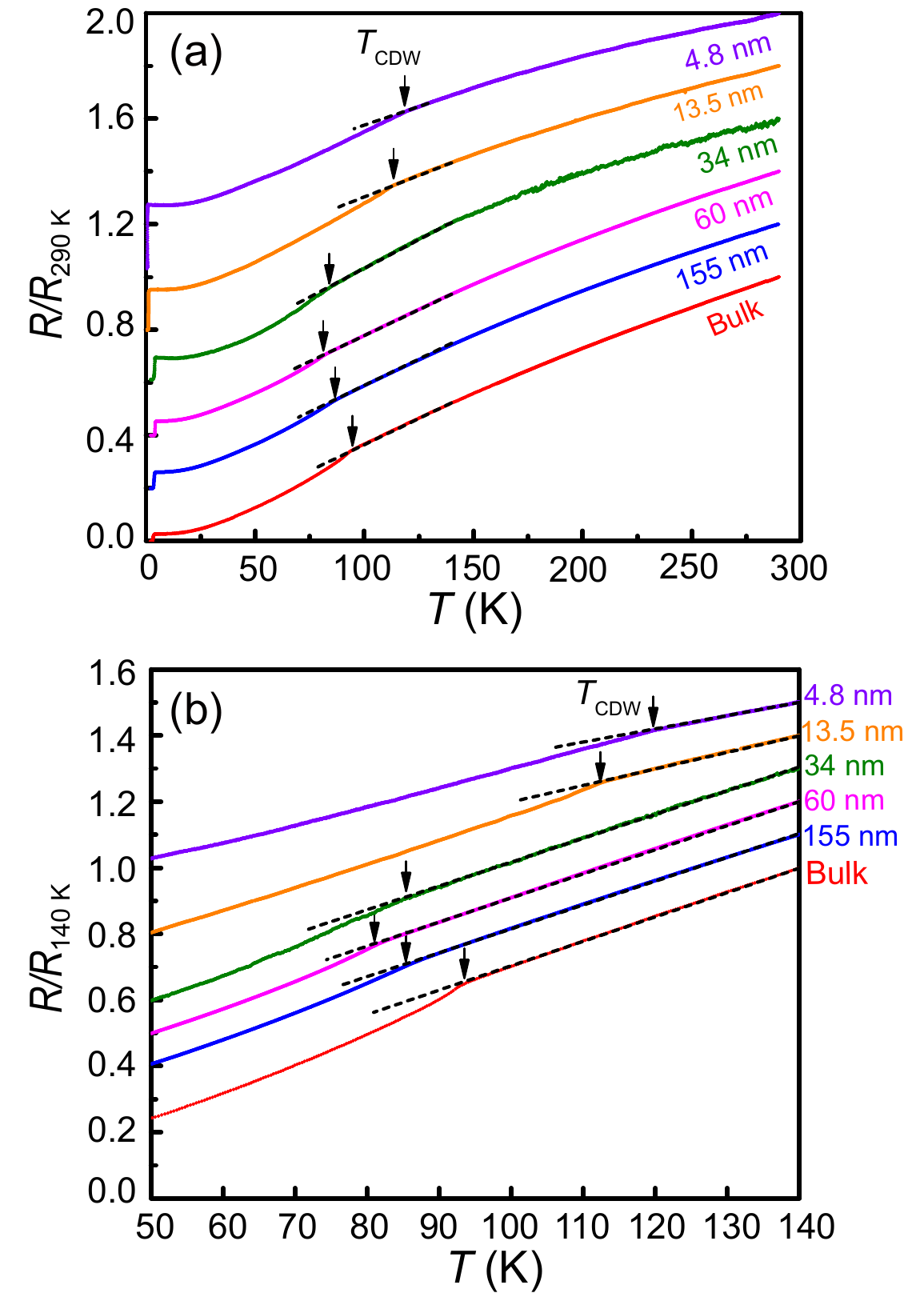}
\caption{Temperature dependence of the resistance for CsV$_3$Sb$_5$ thin flakes with various thickness, from bulk single crystal to 4.8 nm. The data are normalized by the values at 290 and 140 K for (a) and (b), respectively. The curves are vertically shifted for clarity. The arrows mark the CDW transition temperature ${\it T}_{\rm CDW}$. For the bulk single crystal, ${\it T}_{\rm CDW}$ $\approx$ 94 K is consistent with previous report \cite{CsV3Sb5-Z2}.}
\end{figure}

As plotted in Fig. 1(a), CsV$_3$Sb$_5$ consists an ideal Kagome lattice of vanadium coordinated by antimony, with the alkali-metal atoms of Cs intercalated between each layer. The lattice parameters are $a$ = $b$ = 5.4949(3) \AA\ and $c$ = 9.3085(5) \AA\ \cite{Rb-K-Cs}. The single crystal can be easily exfoliated to tens of nanometers by using conventional scotch tape and PDMS films, however such method is difficult to obtain even thinner flakes of CsV$_3$Sb$_5$ sample. With the help of Al$_2$O$_3$, high-quality CsV$_3$Sb$_5$ single crystal can be exfoliated to a few nanometers so that our study can go down to several monolayers. A typical optical image of a CsV$_3$Sb$_5$ thin flake device is shown in Fig. 1(b). Transparent Al$_2$O$_3$ on SiO$_2$/Si substrate manifests green because of diffraction. The yellow part on the Al$_2$O$_3$ is the sample area. A standard four-probe method is used in the resistance measurement. The schematic device structure is plotted in Fig. 1(c). The thickness of the sample was determined by AFM in the selected area marked by red box in Fig. 1(b), along the white line which crosses the edge of the sample in AFM image shown in Fig. 1(d). The AFM measurement gives the thickness of this flake is about 4.8 nm (Fig. 1(e)), corresponding to five monolayers.

Temperature dependence of the resistance for CsV$_3$Sb$_5$ thin flakes with various thickness, from bulk single crystal to 4.8 nm, are shown in Fig. 2. The data are normalized by the values at 290 and 140 K for (a) and (b), respectively. The curves are vertically shifted for clarity. The arrows mark the CDW transition temperature ${\it T}_{\rm CDW}$. For the bulk single crystal, ${\it T}_{\rm CDW}$ $\approx$ 94 K is consistent with previous report \cite{CsV3Sb5-Z2}. With decreasing of the thickness, ${\it T}_{\rm CDW}$ decreases from 94 K in bulk to a minimum of 82 K at thickness of 60 nm, then increases to 120 K as the thickness is further reduced to 4.8 nm (about five monolayers). Such a non-monotonic evolution of ${\it T}_{\rm CDW}$ with sample thickness in CsV$_3$Sb$_5$ is quite unusual. To our knowledge, only VSe$_2$ was reported to manifest this kind of non-monotonic thickness dependent of ${\it T}_{\rm CDW}$ \cite {VSe2}. It was attributed to the crossover in the Fermi surface topology from 3D to 2D around 20 nm \cite {VSe2}. For bulk CsV$_3$Sb$_5$, 2 $\times$ 2 $\times$ 2 superstructure \cite{2*2*2} and 2 $\times$ 2 $\times$ 4 \cite{Fermi surface mapping} superstructure were reported, by means of X-ray diffraction, suggesting 3D CDW. Therefore, the non-monotonic evolution of ${\it T}_{\rm CDW}$ with sample thickness in CsV$_3$Sb$_5$ may be also due to a crossover from 3D to 2D around 60 nm. Note that 60 nm is actually quite thick, far from the 2D limit.

\begin{figure}
\includegraphics[clip,width=7.5cm]{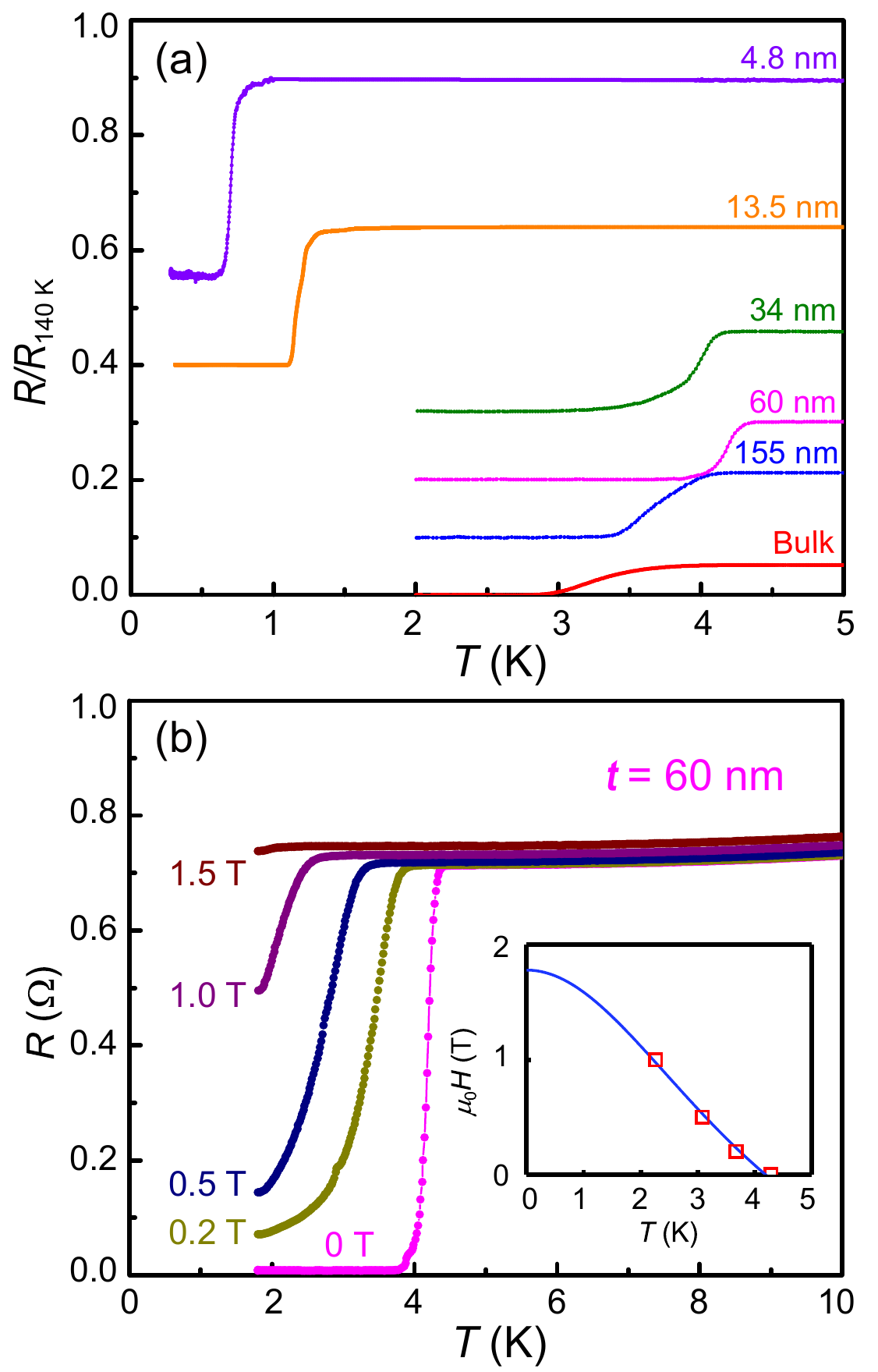}
\caption{(a) The low-temperature resistance for CsV$_3$Sb$_5$ thin flakes with various thickness, normalized by the values at 140 K. The curves are vertically shifted for clarity. (b) Superconducting transitions of the 60 nm CsV$_3$Sb$_5$ flake under various perpendicular magnetic fields up to 1.5 T. The inset shows temperature dependence of the upper critical field $\mu$$_0$${\it H}_{\rm c2}$. Red line marks the fitting of Ginzburg-Landau theory.}
\end{figure}

Figure 3(a) plots the low-temperature resistance below 5 K, to show the superconducting transitions. The superconducting transition temperature ${\it T}_{\rm c}$ is defined as the 10\% drop of the normal-state resistance. It is found that the ${\it T}_{\rm c}$ first increases from 3.64 K in bulk to a maximum of 4.28 K in the 60 nm sample, then decreases to 0.76 K in the 4.8 nm sample. To confirm that the resistance drop in the 60 nm sample is an enhanced superconducting transition, various magnetic fields are applied, as seen in Fig. 3(b). The resistance drop is suppressed with increasing magnetic field, confirming that it is a superconducting transition. The temperature dependence of upper critical field $\mu$$_0$${\it H}_{\rm c2}$ is plotted in the insets of Fig. 3(b). The data can be well fitted by the Ginzburg-Landau (GL) formula $\mu$$_0$${\it H}_{\rm c2}$($T$)= $\mu$$_0$${\it H}_{\rm c2}$(0)(1-($T$/${\it T}_{\rm c}$)$^2$)/(1+($T$/${\it T}_{\rm c}$)$^2$), giving $\mu$$_0$${\it H}_{\rm c2}$(0) $\approx$ 1.78 T. This value is much higher than that of bulk single crystal \cite{SYLiCsVSb}. It is worthy to note that the superconducting transition become sharper with the decreasing of the sample thickness.

Based on above resistance measurements of CsV$_3$Sb$_5$ thin flakes, the evolutions of ${\it T}_{\rm CDW}$ and ${\it T}_{\rm c}$ are plotted in Fig. 4. One can see that ${\it T}_{\rm c}$ shows an exactly opposite evolution, compared with that of ${\it T}_{\rm CDW}$. Such exactly opposite evolutions provide strong evidence for competing superconductivity and CDW in CsV$_3$Sb$_5$.

\begin{figure}
\includegraphics[clip,width=8.5cm]{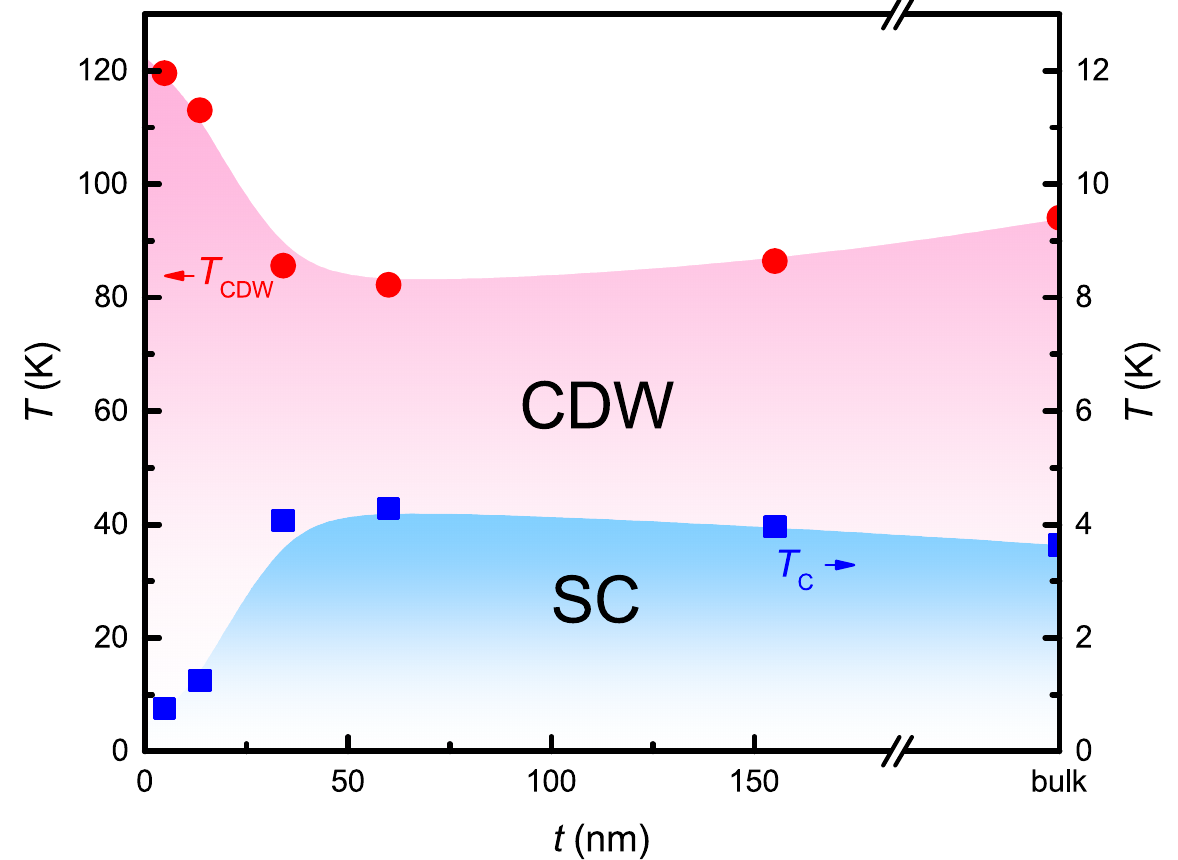}
\caption{The evolutions of superconductivity and charge-density wave with sample thickness in CsV$_3$Sb$_5$. The charge-density wave transition temperature ${\it T}_{\rm CDW}$ decreases first, then increases as the sample thickness is further reduced. Such a non-monotonic evolution can be explained by a 3D to 2D crossover around 60 nm. At the mean time, the superconducting transition temperature $T_c$ of CsV$_3$Sb$_5$ shows an exactly opposite evolution with sample thickness. This provides strong evidence for competing superconductivity and charge-density wave in CsV$_3$Sb$_5$.}
\end{figure}

In summary, we investigate the dimensionality effect on superconductivity and CDW of the new Kagome metal CsV$_3$Sb$_5$ by electrical transport measurements. The opposite non-monotonic evolutions of superconductivity and CDW with the sample thickness give strong evidence for competing superconductivity and CDW. The non-monotonic evolution of ${\it T}_{\rm CDW}$ with reducing sample thickness can be explained by a 3D to 2D crossover around 60 nm. More theoretical calculations and experimental works are needed to clarify the underlying physics of this kind of competition.\\

This work was supported by the Natural Science Foundation of China (Grant No. 12034004), the Ministry of Science and Technology of China (Grant No.: 2016YFA0300503), and the Shanghai Municipal Science and Technology Major Project (Grant No. 2019SHZDZX01). Y. F. Guo was supported by the Major Research Plan of the National Natural Science Foundation of China (No. 92065201) and the Program for Professor of Special Appointment (Shanghai Eastern Scholar). H. C. Lei was supported by National Natural Science Foundation of China (Grant No. 11822412 and 11774423), the Ministry of Science and Technology of China (Grant No. 2018YFE0202600 and 2016YFA0300504), and Beijing Natural Science Foundation (Grant No. Z200005).

B. Q. Song, X. M. Kong, W. Xia, and Q. W. Yin contributed equally to this work.

\noindent $^\dag$ E-mail: hlei$@$ruc.edu.cn\\
\noindent $^\ddag$ E-mail: guoyf$@$shanghaitech.edu.cn\\
\noindent $^\sharp$ E-mail: yangxiaofan$@$fudan.edu.cn\\
\noindent $^*$ E-mail: shiyan$\_$li$@$fudan.edu.cn


\begin{thebibliography}{99}


\bibitem{Rb-K-Cs} B. R. Ortiz, L. C. Gomes, J. R. Morey, M. Winiarski, M. Bordelon, J. S. Mangum, I. W. H. Oswald, J. A. Rodriguez-Rivera, J. R. Neilson, S. D. Wilson, E. Ertekin, T. M. McQueen, and E. S. Toberer, New Kagome prototype materials: discovery of KV$_3$Sb$_5$, RbV$_3$Sb$_5$, and CsV$_3$Sb$_5$. Phys. Rev. Mater. {\bf 3}, 094407 (2019).
\bibitem{KVSb-Z2} E. M. Kenney, M. J. Graf, S. M. L. Teicher, R. Seshadri, and S. D. Wilson, Superconductivity in $Z$$_2$ Kagome metal KV$_3$Sb$_5$, Phys. Rev. Mater. {\bf 5} 034801 (2021).
\bibitem{CsV3Sb5-Z2} B. R. Ortiz, S. M. L. Teicher, Y. Hu, J. L. Zuo, P. M. Sarte, E. C. Schueller, A. M. M. Abeykoon, M. J. Krogstad, S. Rosenkranz, R. Osbron, R. Seshadri, L. Balent, J. He, and S. D. Wilson, CsV$_3$Sb$_5$: A $Z$$_2$ topological Kagome metal with a superconducting ground state. Phys. Rev. Lett. {\bf 125}, 247002 (2020).
\bibitem{RbVSb-SC} Q. W. Yin, Z. J. Tu, C. S. Gong, Y. Fu, S. H. Yan, and H. C. Lei, Superconductivity and normal-state properties of Kagome metal RbV$_3$Sb$_5$ single crystal. Chin. Phys. Lett. {\bf 38}, 037403 (2021).
\bibitem{topological_charge_order_KVSb} Y.-X. Jiang, J.-X. Yin, M. M. Denner, N. Shumiya, B. R. Ortiz, J. He, X. Liu, S. S. Zhang, G. Chang, I. Belopolski, Q. Zhang, M. S. Hossain, T. A. Cochran, D. Multer, M. Litskevich, Z.-J. Cheng, X. P. Yang, Z. Guguchia, G. Xu, Z. Wang, T. Neupert, S. D. Wilson, and M. Z. Hasan, Discovery of topological charge order in Kagome superconductor KV$_3$Sb$_5$, arXiv: 2012.15709 (2020).
\bibitem{anomalous Hall effect} S. Y. Yang, Y. Wang, B. R. Ortiz, D. Liu, J. Gayles, E. Derunova, R. Gonzalz-Hernandez, L. Smejkal, Y. Chen, S. S. P. Parkin, S. D. Wilson, E. S. Toberer, T. McQueen, M. N. Ali, Giant, unconventional anomalous Hall effect in the metallic frustrated magnet candidate, KV$_3$Sb$_5$, Sci. Adv. {\bf 6}, 6003 (2020).
\bibitem{SYLiCsVSb} C. C. Zhao, L. S. Wang, W. Xia, Q. W. Yin, J. M. Ni, Y. Y. Huang, C. P. Tu, Z. C. Tao, Z. J. Tu, C. S. Gong, H. C. Lei, Y. F. Guo, X. F. Yang, and S. Y. Li, Unconventional superconductivity in the topological Kagome metal CsV$_3$Sb$_5$, arXiv: 2102.08356 (2021).
\bibitem{Strong-coupling} H. Chen, H. Yang, B. Hu, Z. Zhao, J. Yuan, Y. Xing, G. Qian, Z. Huang, G. Li, Y. Ye, Q. Yin, C. Gong, Z. Tu, H. Lei, S. Ma, H. Zhang, S. Ni, H. Tan, C. Shen, X. Dong, B. Yan, Z. Wang, and H.-J. Gao, Roton pair density wave and unconventional strong-coupling superconductivity in a topological kagome metal, arXiv: 2103.09188.
\bibitem{Nodeless superconductivity} W. Duan, Z. Nie, S. Luo, F. Yu, B. R. Ortiz, L. Yin, H. Su, F. Du, A. Wang, Y. Chen, X. Lu, J. Ying, S. D. Wilson, X. Chen, Y. Song, and H. Yuan, Nodeless superconductivity in the kagome metal CsV$_3$Sb$_5$. arXiv: 2103.11796 (2021).
\bibitem{s-wave superconductivity} C. Mu, Q. Yin, Z. Tu, C. Gong, H. Lei, Z. Li, and J. Luo, $s$-wave superconductivity in kagome metal CsV$_3$Sb$_5$ revealed by $^{121/123}$Sb NQR and $^{51}$V NMR measurements. arXiv: 2104.06698 (2021).
\bibitem{Multiband superconductivity} H.-S. Xu, Y.-J. Yan, R. Yin, W. Xia, S. Fang, Z. Chen, Y. Li, W. Yang, Y. Guo, and D.-L. Feng, Multiband superconductivity with sign-preserving order parameter in kagome superconductor CsV$_3$Sb$_5$. arXiv: 2104.08810 (2021).
\bibitem{Fermi surface mapping}B. R. Ortiz, S. M. L. Teicher, L. Kautzch, P. M. Sarte, J. P. C. Ruff, R. Seshadri, and S. D. Wilson, Fermi surface mapping and the nature of charge density wave order in the kagome superconductor CsV$_3$Sb$_5$. arXiv: 2104.07230 (2021).
\bibitem{J.-G. Cheng}  K. Y. Chen, N. N. Wang, Q. W. Yin, Z. J. Tu, C. S. Gong, J. P. Sun, H. C. Lei, Y. Uwatoko, and J.-G. Cheng, Double superconducting dome and triple enhancement of ${\it T}_{\rm c}$ in the Kagome superconductor CsV$_3$Sb$_5$ under high pressure. arXiv: 2102.09328 (2021).
\bibitem{Z. Yang} Z. Zhang, Z. Chen, Y. Zhou, Y. Yuan, S. Wang, L. Zhang, X. Zhu, Y. Zhou, X. Chen, J. Zhou, and Z. Yang, Pressure-induced reemergence of superconductivity in topological Kagome metal CsV$_3$Sb$_5$, arXiv: 2103.12507 (2021).
\bibitem{X. Chen}  X. Chen, X. Zhan, X. Wang, J. Deng, X.-B. Liu, X. Chen, J.-G. Guo, X. Chen, Highly-robust reentrant superconductivity in CsV$_3$Sb$_5$ under pressure. arXiv: 2103.13759 (2021).
\bibitem{Pressure-KVSb}F. Du, S. Luo, B. R. Ortiz, Y. Chen, W. Duan, D. Zhang, X. Lu, S. D. Wilson, Y. Song, and H. Yuan, Pressure-tuned interplay between charge order and superconductivity in the Kagome metal KV$_3$Sb$_5$. arXiv: 2102.10959v1 (2021).
\bibitem{Al2O3}Y. Deng, Y. Yu, Y. Song, J. Zhang, N. Z. Wang, Z. Sun, Y. Yi, Y. Z. Wu, S. Wu, J. Zhu, J. Wang, X. H. Chen, and Y. Zhang, Gate-tunable room-temperature ferromagnetism in two-dimensional Fe$_3$GeTe$_2$, Nature {\bf 563}, 94 (2018).
\bibitem{VSe2} \'{A}. P\'{a}sztor, A. Scarfato, C. Barreteau, E. Giannini, and C. Renner, Dimensional crossover of the charge density wave transition in thin exfoliated VSe$_2$, 2D Mater. {\bf 4} 041005 (2017).
\bibitem{2*2*2} H. X. Li, T. T. Zhang, Y.-Y. Pai, C. Marvinney, A. Said, T. Yilmza, Q. Yin, C. Gong, Z. Tu, E. Vescovo, R. G. Moore, S. Murakami, H. C. Lei, H. N. Lee, B. Lawrie, and H. Miao, Observation of unconventional charge density wave without acoustic phonon anomaly in Kagome superconductors AV$_3$Sb$_5$ (A = Rb,Cs), arXiv:2103.09769 (2021).



\end{thebibliography}
\end{document}